# Spin orbit interaction induced spin-separation in platinum nanostructures

Koong Chee Weng[1,2], N. Chandrasekhar[1,2], Christian Miniatura[1,3]
Berthold-Georg Englert[1,4]

[1]Department of Physics, National University of Singapore, Singapore 117542
[2] IMRE, 3 Research Link, Singapore 117602
[3] INLN, UNS, CNRS, 1361 route des Lucioles, F-06560 Valbonne, France
[4] Centre for Quantum Technologies, National University of Singapore, Singapore 117543

**Abstract**

Hirsch (1999) proposed a mechanism and geometry for the observation of the spin-Hall effect. In this work, we present a novel realization of the Hirsch geometry in a platinum (Pt) nanostructure, which is an increasingly important material for spintronics applications. Measurements were made in a non-local geometry to avoid spurious effects. The measurements show the large spin Hall conductivity of Pt. The results are compared with gold (Au) and aluminum (Al). Possible theoretical explanations of our observations are briefly mentioned.

## 1 Introduction

Spintronics, a subfield of condensed matter physics, promises a new generation of electronic devices, including reconfigurable device architectures [1]. Conventional electronic devices use only the charge state of an electron, whereas spintronics exploits the electron's spin degree of freedom to carry, manipulate and store information. Spintronics requires an understanding of spin injection, manipulation and detection. Typical spintronic devices (e.g. spin valve) use a ferromagnetic metal to inject spin polarized carriers into another metal [2]. However spin polarized carrier injection into semiconductors is fraught with problems due to the interface between the electronically and structurally dissimilar materials [3]. Therefore, recent work has been targeted towards understanding and exploiting spin polarized currents by intrinsic mechanisms in a single material. Once such mechanism is the spin Hall effect (SHE) [4], which refers to the generation of a transverse spin current in materials with spin-orbit interaction due to an applied longitudinal electrical field. For a conductor of finite size, this causes the electrons with different spin states to accumulate at opposite sides.

The development in semiconductor SHE has limited applications due to the small output and low temperature requirement. However, recent work in paramagnetic metal

SHE has shown possible room temperature SHE [5, 6]. Interestingly the metallic spin Hall conductivity effect at room temperature is larger than the semiconductor spin Hall conductivity at 4 K [5]. As present spintronics applications rely heavily on the use of ferromagnetic metals as either a spin detector (e.g. hard drive read head) or as a memory storage (e.g. magnetoresistive random access memory), the smaller conductivity mismatch between paramagnetic metals and ferromagnetic metals allows for effective integration of SHE with ferromagnets to improve present technology or to develop novel applications.

However, from practical consideration, such spin currents are impossible to detect directly by conventional electronic means. Currently the measurement of SHE is done via optical means or by using a ferromagnetic metal as spin detector. In 1999 Hirsch [7] proposed a device geometry for the electrical detection of such spin separation. But, this geometry had attracted little or no experimental interest due to the practical difficulties in fabrication and measurement. In this work, we present a novel realization of the Hirsch geometry using conventional device fabrication techniques. Measurements were made in a non-local geometry to remove spurious contributions to the readings. Measurement results for Pt are contrasted with results for Au and Al for a qualitative comparison. Possible theoretical explanations of our observations are briefly mentioned.

## 2 Spin-Orbit Interaction and Its Effect on Electron Spin

Spin-orbit interaction, as a mechanism of spin relaxation or dephasing was first predicted by D'yakonov and Perel' [8]. This mechanism applies to bulk III-V and two dimensional III-V structures. The situation in paramagnetic metals, on the other hand, is completely different. Hirsch [7] was the first to postulate that either skew spin scattering by impurities, or strong spin orbit interaction could generate a spin current in paramagnetic metals. This has been termed the spin Hall effect (SHE), and has not been observed until recently. SHE did not attract much attention until Murakami *et al.* [9] predicted a dissipationless spin current that could originate from the band structure, and since sparked extensive theoretical and experimental studies [10, 11, 12, 5, 13, 14, 15, 6] . The arguments of Murakami *et al.* are based on analogies with the quantum Hall effect (QHE), and apply in particular to the split off heavy hole band in III-V semiconductors. Rashba [16] has pointed out that such spin currents, caused by spin orbit interaction, are not transport currents which could be employed for transporting spins or spin injection, in thermodynamic equilibrium. Subsequently two landmark experiments have reported the observation of spin accumulation. In n-doped bulk GaAs and InGaAs [10], the observation was attributed to skew scattering of spins by impurities [17], referred to as the extrinsic effect which also included the side-jump mechanism [18]. However, observations on spin accumulation of a two-dimensional hole gas (2DHG)in a p-doped GaAs structure [11] were attributed to intrinsic effects, such as the mechanism proposed by Murakami *et al.* [9].

More recently, there has been significant experimental development in SHE and inverse SHE (spin current producing charge accumulation) in metallic system. Valenzuela and Tinkham [12] have reported the first electrical detection of the SHE in alu-minum strips at 4.2 K. In their experiment, a spin polarized charge current is injected from a ferromagnetic electrode into the Al strips and the pure spin current is "filtered" out using a non-local measurement geometry. Similarly, Kimura *et al.* [5] generated and detected spin accumulation in Pt using lateral ferromagnetic-nonmagnetic (FM/NM) devices at room temperature (RT) and 77 K.

Spurred by these observations, first-principles calculations by Yao and Fang [13], and Guo *et al.* [14], have shown that simple metals like tungsten, gold and platinum can have larger intrinsic spin Hall conductivity and are more robust against disorder compared to semiconductors such as GaAs and Si. The calculated intrinsic SHE is insensitive to disorder [13, 14]. Their calculations are based on standard density functional theory using accurate full potential linearized augmented plane wave (FLAPW) and full-potential linear-muffin-tin-orbital (FPLMTO) methods respectively. The relativistic spin-orbit coupling has been treated fully self-consistently and the intrinsic spin Hall conductivity is calculated by using a generalized Kubo formula. The intrinsic effect is interpreted as the integral of the Berry phase curvature, $\mathbf{\Omega}(\mathbf{k})$ over the occupied electronic states, which acts as an effectively spin dependent magnetic field in crystal momentum space to produce a transverse velocity proportional to $\mathbf{E} \times \mathbf{\Omega}(\mathbf{k})$ [19, 20, 21] (Refer to [22] for a comprehensive review of Berry phase). Several other related studies in anomalous Hall effect (AHE) have demonstrated the validity of such an approach [23]. Other numerical studies have provided numerical evidence that the intrinsic spin Hall effect can survive weak disorder in the mesoscopic diffusive regime [15, 24]. It must be stressed that, while the precise physical mechanism for the spin-orbit interaction induced spin separation in semiconductor has reached general consensus among the community, the spin separation in paramagnetic metals remains unknown or obscure at this point. Despite these controversial results, the various advancements mentioned above spurred widespread interest in the SHE in semiconductor and metallic structures.

The prospect of using metallic SHE based spin devices is a potentially attractive alternative to the current paradigm. Currently a ferromagnet (FM) is the *de facto* spin source and/or spin detector in the study of metallic (NM) spin transport. However the use of FM as spin source poses a problem, namely the conductivity mismatch at the FM/NM interface which disrupts the transfer of spin polarization across the interface [3]. Furthermore the position of the FM spin detector is required to be within the NM spin diffusion length, as measured from the spin source. It is evident that FM/NM based spin device design would be spatially restricted by the NM spin diffusion length which is of the order of a few tens of nanometers and depends on the operating temperature. In contrast, the requirement of a minimal shielding distance would limit the proximity of the neighboring magnetic electrodes. Additionally, the ferromagnet based spin devices require magnetic control of multiple ferromagnets, thereby posing a difficult problem. In our work, we propose an alternative method of testing the metallic SHE without the use of ferromagnet. As such, our scheme circumvents the FM/NM spin injection problems and also avoids other spurious magnetoresistance effects (e.g. anisotropic magnetoresistance and anomalous Hall effect).

# 3 Experimental Procedure

The spin Hall effect postulates a force which is perpendicular to both velocity and spin orientation for carriers moving in a conductor with non-zero spin-orbit interaction. Clearly, in accordance with Onsager's reciprocal relations [25], the same spin dependent force would cause the electrons in a pure spin current to be deflected to one side, thereby generating a transverse charge current. This "Hall-like" behavior results in charge accumulation at the sides, enabling a voltage signal to be measured as a signature of what was initially a pure spin current. This is the essential physics of the device proposed by Hirsch. In order to exploit the reciprocal relation governing the conversion of the spin current to a charge current and the associated potential difference, severe geometrical constraints have to be imposed on the device configuration. This entails the fabrication of quantum point contacts at opposite transverse locations on the sample, laying down an insulating layer, and having the same paramagnetic metal as the third layer, which is electrically connected to the quantum point contacts (see Hirsch's paper for a complete description [7]). The potential difference is manifested in the longitudinal direction of this top third layer. In our work, we design and test a non-magnetic lateral structure based on the principle of SHE, and successfully demonstrate the electrical generation and detection of electron spin in a platinum nanostructure.

We have adopted a two dimensional version of Hirsch's experimental geometry. Our single layer design avoids the fabrication complexity of Hirsch's multilayer structure while maintaining the essence of his proposal. A similar geometry was proposed by Hankiewicz *et al.* to measure the spin Hall conductivity in mesoscopic-scale system [15]. We create and detect electron spin using the SHE, in order to measure the spin Hall conductivity and the spin diffusion length with a non-local four-terminal setup. The device consists of three parallel conductors with a perpendicular bridging conductor. The devices are prepared with conventional e-beam lithography techniques, DC sputtering and lift-off process. The Pt wire is 200 nm wide with a thickness of 50 nm. The separation, $L_{sH}$, between each Pt channel varied from 180 to 320 nm (Figure 1a). The current-voltage (I-V) data are measured in a cryostat using a Keithley sourcemeter and nanovoltmeter. The experiment is also repeated for Au and Al in order to compare the SHE in metals with different spin-orbit interaction strength.

The reservoirs A and B are biased to produce a current in the *x* direction. The distance between A and B is large such that we can assume that the current density is uniform longitudinally. Due to spin-orbit interaction, a transverse spin current is created and diffuses through the bridging conductor in our experimental configuration (see Figure 1a). Then the pure spin current would again be deflected due to the spin-orbit interaction and lead to a charge accumulation on the sides of the bridging conductor. To put it simply, a charge current is passed between electrode A and B, i.e. $I_{BA}$ and leads to a pure spin current, $j_s$, in the *y* direction. Subsequently, the spin current leads to a charge accumulation which is measured as a voltage between electrode C and D.

Following the analysis presented by Adagideli *et al.* [26], we assume that the charge and spin transport is diffusive and that the spin current, $j_{y,max}^z$, reaches its maximum value at the interface at region 2 and 3 shown in Figure 1a. We can write down the expression of the spin current along the transverse or bridging conductor as

$$j_s(y') = j_{y,max}^z \exp\left(\frac{-y'}{\lambda_{sf}}\right) \qquad \text{where} \qquad y' = \begin{cases} y - \frac{w}{2} : y \geq \frac{w}{2} \\ y + \frac{w}{2} : y \leq \frac{w}{2} \end{cases} \qquad (1)$$

where $\lambda_{sf}$ is the spin diffusion length, $w$ is the centre conductor width and $y$ is the lateral distance measured from the centre axis.

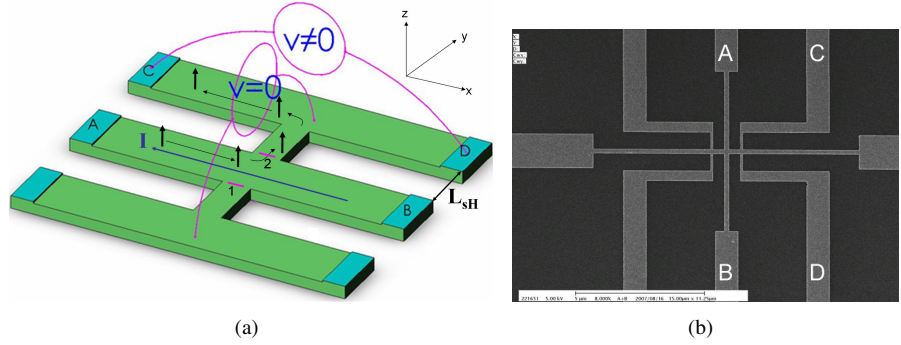

(a) (b)

Figure 1: The spin Hall effect and the inverse spin Hall effect. (a) When $I_{BA}$ is applied, the traversing spin-down (spin-up) electrons are deflected to the left (right) which lead to spin current in the $y$ direction. The spin current diffuses along the bridging conductor (transverse channel), and these carriers are again deflected in accordance with the Onsager relations. The resulting effect of the spin imbalance is a spin current induced a charge accumulation between the measurement probes C and D. The signal is termed the spin Hall voltage, $V_{sH}$. (b) Scanning electron microscopy image of a representative device.

In general, the charge current in a conductor is [27]

$$\mathbf{j}_c(\mathbf{r}) = \sigma \mathbf{E}(\mathbf{r}) + \frac{\sigma_{sH}}{\sigma}(\hat{\mathbf{z}} \times \mathbf{j}_s), \qquad (2)$$

where the first term arises from the applied electrical field and the second term from the inverse spin Hall effect (ISHE), where we only consider the spin polarization in the $z$ direction. ISHE is the conversion of a spin current into an electric current as a consequence of Onsager's reciprocal relations between spin current and charge current. $\sigma_{sH}$ is the spin Hall conductivity and $\sigma$ is the Drude conductivity.

By substituting $j_s$ from Equation (1) into (2), and setting $E(y) = 0$ we have

$$E_x(y') = -\frac{\sigma_{sH}}{\sigma^2} j_s(y') = -\frac{\sigma_{sH}}{\sigma^2} j_{y,max}^z \exp\left(\frac{-y'}{\lambda_{sf}}\right). \qquad (3)$$

At $y' = L_{sH}$, the spin Hall voltage $V_{sH}$ measured using electrode C and D is:

$$V_{sH} = -dE_x(L_{sH}) = d\frac{\sigma_{sH}}{\sigma^2} j^z_{y,max} \exp\left(\frac{-L_{sH}}{\lambda_{sf}}\right), \tag{4}$$

where $d$ is the width of the transverse conductor.

The spin Hall resistance $R_{sH} = V_{sH}/I$ is

$$\ln R_{sH} = \left(\frac{-L_{sH}}{\lambda_{sf}}\right) + \ln\left(d\frac{\sigma_{sH}}{\sigma^2}\frac{j^z_{y,max}}{I}\right). \tag{5}$$

By measuring $\ln R_{sH}$ for various $L_{sH}$, the value of $\lambda_{sf}$ can be extracted using Equation (5). However the value of $\sigma_{sH}$ and $j^z_{y,max}$ could not be individually determined, i.e. only the product of the two quantities can be calculated. However, the value of $\sigma_{sH}$ can be obtained using optical methods similar to the Kerr rotation method used by Kato *et al* [10]. The optical work will be explained in a later publication. Given the value of $\sigma_{sH}$, the value of $j^z_{y,max}$ can be calculated.

## 4 Results and Discussion

Figure 2 shows representative V-I results for our devices for $L_{sH}$ = 237 nm. At room temperature the measured voltage does not show any discernible dependence on the current. At lower temperature, a linear relationship appears.

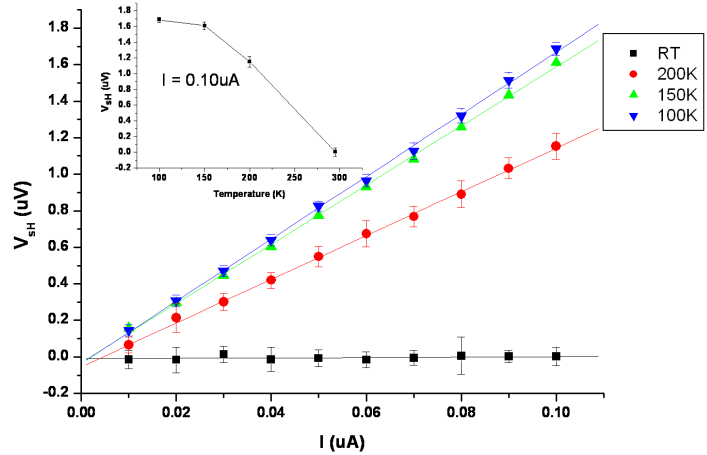

Figure 2: $V_{sH}$ as a function of current I for various temperature. Inset, $V_{sH}$ as a function of temperature for $I = 0.1\mu A$. RT = 298K.

Before we discuss these results, some comments are in order. As a consequence of our experimental geometry, the measured voltage cannot arise due to any misalignment or defects in our structures, nor can it arise due to any magnetic field, since no external field was switched on. The influence of stray fields can be ignored, since Hall probe measurements in the vicinity of the measurement region yield fields of the order of the earth's magnetic field. These results clearly indicate a voltage that scales with the current.

Figure 3 shows the measured value of $R_{sH}$ as a function of the distance, $L_{sH}$, between the two channel. Consistent with Equation (4), $R_{sH}$ decreases as a function of $L_{sH}$. By linear fitting of the graph to Equation (4) the value of $\lambda_{sf}$ and $\sigma_{sH} j^z_{y,max}$ are obtained. Using the value of $\sigma$ measured from our sample ($4.06 \times 10^6$ $\Omega^{-1}\text{m}^{-1}$ at 100 K), the value of $\lambda_{sf}$ is found to be 16 nm. Our calculated value of $\lambda_{sf}$ is of the same order of magnitude as reported elsewhere [28] (14 nm at 4 K). Similarly the value of $\sigma_{sH} j^z_{y,max}$ is found to be $8.9 \times 10^{-9}$ $\Omega^{-1}\text{m}^{-3}$A. We take the literature value of $2.4 \times 10^4$ $\Omega^{-1}\text{m}^{-1}$ at RT [5] as a lower limit estimate for $\sigma_{sH}$, and we get $j^z_{y,max}$ to be $3.7 \times 10^{-15}$ Am$^{-2}$. This sets the lower limit for the spin current $j_s$ for our devices.

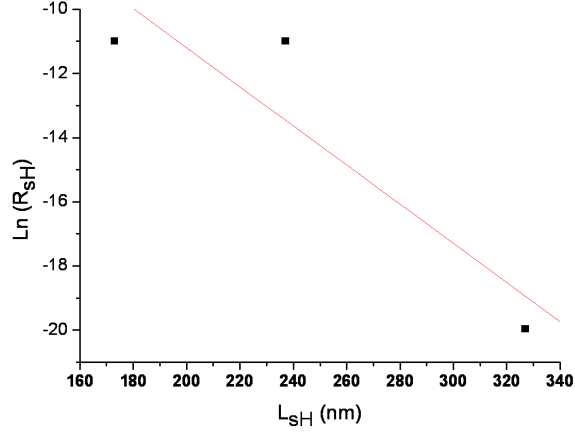

Figure 3: $R_{sH}$ as a function of $L_{sH}$.

The experimental results for Au and Al are shown in Figure 4. The signal for Au shows a linear response similar to Pt, while the signal for Al shows no noticeable signal above instrumental noise. This is not expected as Al, a low atomic number metal, has a weak spin-orbit interaction and Au, a high atomic number metal, should have stronger spin-orbit interaction since the spin-orbit interaction is expected to scale with the fourth power of the atomic number [29]. However, the signal difference, $R_{sH}$, between Pt (~ 10Ω) and Au (~ 0.02Ω) is huge even though the atomic numbers of Au (79) and Pt (78) are similar. If we consider the SHE to be intrinsic in nature, where the band structure plays an important role, the unconventional result can be explained in the following context. As the band structure near the fermi level for Au is mainly S-type while for Pt is mainly D-type [14], it is predicted that the intrinsic SHE is larger for Pt than Au. However further experimental studies need to be conducted to confirm this hypothesis.

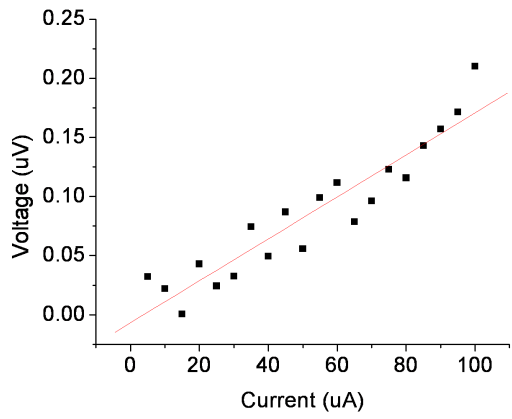 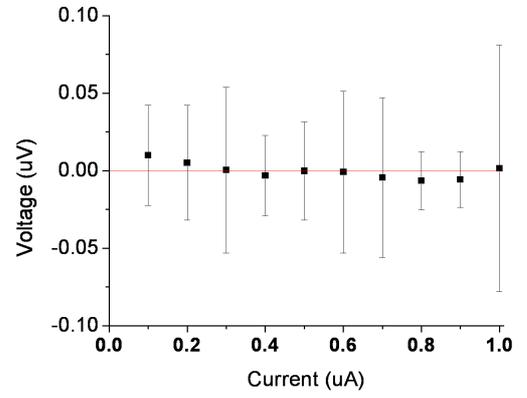

(a) (b)

Figure 4: The current-voltage measurement for (a) Au and (b) Al.

# 5 Conclusion

We have demonstrated the electrical generation and detection of spin via the SHE and inverse SHE in Pt. We also compared the signal of Pt with Au and Al. The spin diffusion length without the use of ferromagnetic material. The numbers are found to be consistent with other reports in the literature. Spin orbit interaction can be a useful method to induce spin separation in metallic nanostructures of high atomic number metals. Whether this spin separation can be put to any practical use, remains to be seen.

We thank G. Y. Guo for useful discussions. This work was supported by A*Star (Grants No. 012-104-0040) and by National University of Singapore (Grants No. WBS:R-144-000-179-112).

# References

[1] Žutić, I., Fabian, J., and Das Sarma, S. *Reviews of Modern Physics* **76**, 323 (2004).

[2] Jedema, F. J., Filip, A. T., and van Wees, B. J. *Nature* **410**, 345–348 (2001).

[3] Filip, A. T., Jedema, F. J., van Wees, B. J., and Borghs, G. *Physica E: Low-Dimensional Systems And Nanostructures* **10**, 478–483 (2001).

[4] Glazov, M. and Kavokin, A. *Journal Of Luminescence* **125**, 118–125 (2007).

[5] Kimura, T., Otani, Y., Sato, T., Takahashi, S., and Maekawa, S. *Physical Review Letters* **98**, 156601 (2007).

[6] Seki, T., Hasegawa, Y., Mitani, S., Takahashi, S., Imamura, H., Nitta, S. M. J., and Takanashi, K. *Nature Materials* **7**, 125–129 (2008).

[7] Hirsch, J. E. *Physical Review Letters* **83**, 1834–1837 (1999).

[8] D'yakonov, M. I. and Perel', V. I. *Physics Letters A* **35**, 459–460 (1971).

[9] Murakami, S., Nagaosa, N., and Zhang, S.-C. *Science* **301**, 1348–1351 (2003).

[10] Kato, Y. K., Myers, R. C., Gossard, A. C., and Awschalom, D. D. *Science* **306**, 1910–1913 (2004).

[11] Wunderlich, J., Kaestner, B., Sinova, J., and Jungwirth, T. *Physical Review Letters* **94**, 047204 (2005).

[12] Valenzuela, S. O. and Tinkham, M. *Nature* **442**, 176–179 (2006).

[13] Yao, Y. and Fang, Z. *Physical Review Letters* **95**, 156601 (2005).

[14] Guo, G. Y., Murakami, S., Chen, T.-W., and Nagaosa, N. *cond-mat/07050409* (2007).

[15] Hankiewicz, E. M., Molenkamp, L. W., Jungwirth, T., and Sinova, J. *Physical Review B* **70**, 241301 (2004).

[16] Rashba, E. I. *Physical Review B* **68**, 241315 (2003).

[17] Smit, J. *Physica* **24**, 39–51 (1958).

[18] Berger, L. *Physical Review B* **2**, 4559–4566 (1970).

[19] Onoda, M. and Nagaosa, N. *Physical Review Letters* **90**, 206601 (2003).

[20] Jungwirth, T., Niu, Q., and MacDonald, A. H. *Physical Review Letters* **88**, 207208 (2002).

[21] Taguchi, Y., Oohara, Y., Yoshizawa, H., Nagaosa, N., and Tokura, Y. *Science* **291**, 2573–2576 (2001).

[22] Ong, N. P. and Lee, W.-L. *cond-mat/0508236* (2005).

[23] Yao, Y., Kleinman, L., MacDonald, A. H., Sinova, J., Jungwirth, T., Wang, D.-S., Wang, E., and Niu, Q. *Physical Review Letters* **92**, 037204 (2004).

[24] Nikolić, B. K., Zarbo, L. P., and Souma, S. *Physical Review B* **72**, 075361 (2005).

[25] Onsager, L. *Physical Review* **38**, 2265–2279 (1931).

[26] Adagideli, I. and Bauer, G. E. W. *Physical Review Letters* **95**, 256602 (2005).

[27] Zhang, S. *Physical Review Letters* **85**, 393–396 (2000).

[28] Kurt, H., Loloee, R., Eid, K., W. P. Pratt, J., and Bass, J. *Applied Physics Letters* **81**, 4787–4789 (2002).

[29] Geier, S. and Bergmann, G. *Physical Review Letters* **68**, 2520–2523 (1992).